\documentclass[aps,prd,amsmath,floats,floatfix,
  superscriptaddress,
  nofootinbib,
  notitlepage,
  showpacs]{revtex4-1}
\usepackage{Preamble}
\usepackage{hyperref}

\newcommand{\bea}{\begin{eqnarray}}
\newcommand{\eea}{\end{eqnarray}}

\begin{document}

\definecolor{orange}{rgb}{0.9,0.45,0} 
\definecolor{applegreen}{rgb}{0.055, 0.591, 0.0530}
\newcommand{\JB}[1]{{\textcolor{green}{[JuanB: #1]}}}
\newcommand{\juanc}[1]{{\textcolor{blue}{[JC: #1]}}}
\newcommand{\nico}[1]{{\textcolor{brown}{[nico: #1]}}}
\newcommand{\dario}[1]{{\textcolor{red}{[Dario: #1]}}}

\author{Nicolas Sanchis-Gual}
\affiliation{Departament d'Astronomia i Astrof\'isica, Universitat de Val\`encia, Av. Vicent Andr\'es Estell\'es 19, 46100 Burjassot (Val\`encia), Spain}

\author{Juan Barranco}
\affiliation{Departamento de F\'isica, Divisi\'on de Ciencias e Ingenier\'ias,
Campus Le\'on, Universidad de Guanajuato, C.P. 37150, Le\'on, M\'exico.}

\author{Juan Carlos Degollado}
\affiliation{Instituto de Ciencias F\'isicas, Universidad Nacional Aut\'onoma de M\'exico,
Apartado Postal 48-3, 62251, Cuernavaca, Morelos, M\'exico}

\author{Darío Nuñez}
\affiliation{Instituto de Ciencias Nucleares, 
Universidad Nacional
Aut\'onoma de M\'exico, A.P. 70-543, M\'exico D.F. 04510, M\'exico.}

\title{
\emph{Dark-to-black} super accretion as a mechanism for early supermassive black hole growth
}


\begin{abstract}
The discovery of supermassive black holes with masses $\gtrsim 10^9 M_\odot$ at redshifts $z\gtrsim 10$ challenges conventional formation scenarios based on baryonic accretion and mergers within the first few hundred million years. We propose an alternative channel in which ultralight scalar dark matter undergoes \emph{dark-to-black} conversion via quasi-bound state depletion around black hole seeds. We estimate the accretion rate of the scalar field as a function of the boson mass parameter $\mu$ and the black hole mass $M_{\rm BH}$, and integrate this rate over cosmological timescales. 
Our results show that once a critical value of $\mu M_{\rm BH}$ is reached, scalar field accretion becomes highly efficient, enabling substantial black hole growth even from relatively small initial seed masses. For boson masses $\mu \sim 10^{-19}$--$10^{-16}\,\mathrm{eV}$, black hole seeds of $10^2$--$10^5 M_\odot$ can reach $10^6$--$10^8 M_\odot$ within $\sim 10^8$ yr. This \emph{dark-to-black} mechanism provides a natural pathway for the rapid formation of massive black holes in the early universe, offering a potential probe of the microphysical nature of dark matter.
\end{abstract}

\maketitle

\section{Introduction}
Supermassive black holes (SMBHs), dark matter, and stars are the three fundamental components of galaxies. Their formation, growth, and evolution are highly coupled and require processes that in many cases last for hundreds to thousands of millions of years. However, recent observations reveal that massive galaxies hosting SMBHs were already in place at very high redshifts  quasars at $z\sim7$ for example are known to contain black holes of mass $\sim10^9\,M_\odot$~\cite{maiolino2024small,banados2018800}, and very recently galaxy candidates have been identified out to $z\sim16$ or more by the deep James Webb Space Telescope (JWST) surveys~\cite{atek2023revealing,robertson2024earliest}, suggesting significant structure formation when the Universe was less than a few hundred million years old. These findings strain the standard pathways by which black hole seeds grow via baryonic accretion and mergers within hierarchical structure formation~\cite{valiante2016first}.

Leading explanations proposed to date include:  
(i) massive Population~III stellar seeds, which collapse into black holes of $10^2$–$10^3\,M_\odot$ and may grow rapidly~\cite{haemmerle2018evolution,kroupa2020very};  
(ii) direct collapse scenarios forming supermassive stars ($\sim10^4$–$10^6\,M_\odot$) that collapse into black hole seeds~\cite{johnson2013supermassive,jeon2025physical};  
(iii) super-Eddington accretion episodes, where growth rates exceed the classical Eddington limit under favorable radiative or geometrical conditions~\cite{salpeter1965accretion,trinca2024episodic}; and  
(iv) frequent mergers of smaller black holes in dense environments~\cite{atallah2023growing,gaete2024supermassive,bamber2025evolution,reinoso2025massive}.

Alternative or complementary mechanisms invoking dark matter have also been considered. For example, recent works (e.g.\ \cite{chiu2025boosting,imai2025dark}) investigate how ultralight dark matter or fuzzy/bosonic dark matter can influence black hole growth either by enhancing gas accretion (via deeper gravitational potential wells due to soliton cores) or by direct accretion of dark matter over cosmic time.  

In parallel, bosonic fields (scalar, vector or axion-like) can resonate with black holes or even neutron stars at different mass scales~\cite{arvanitaki2011exploring,herdeiro2023black,lazarte2025gravitational}. One of the best-known resonant phenomena is superradiance ~\cite{brito2015superradiance}. Superradiance occurs for bosonic fields around rotating black holes, for modes with positive azimuthal number 
$m>0$ whose frequency lies below the critical threshold $\omega<m\Omega_H$, where $\Omega_H$
 is the angular velocity of the horizon. Even then, the corresponding growth timescales are generally extremely long, rendering the effect negligible in most regimes. A notable exception arises when the dimensionless parameter $\mu M_{\rm BH}\sim\mathcal{O}(1)$, where $M_{\rm BH}$ is the black hole mass and $\mu$ is the scalar particle mass (in natural units, $c=G=\hbar=1$), for which the instability is significantly enhanced (see~\cite{dolan2013superradiant,herdeiro2014kerr,east2017superradiant,conlon2018can,degollado2018effective}). In this case, there exists a specific black hole mass range where superradiance becomes the dominant process shaping the dynamics of the scalar cloud.

In this paper we propose a related resonant mechanism, but instead of extracting energy via superradiance, the scalar field forms quasi-bound states around the black hole which decay exponentially, i.e.\ an effective {\it superaccretion} of the bosonic field. Massive bosonic dark matter is known to support quasi-bound states in the vicinity of black holes~\cite{detweiler1980klein,barranco2011black,barranco2012schwarzschild,sanchis2015quasistationary,sanchis2015bquasistationary,sanchis2016quasistationary}, whose decay timescales depend sensitively on both the black hole mass and the field mass. The accretion rate depends on whether the Compton wavelength of the boson $\lambda_c \sim \frac{1}{\mu}$ is greater than the black hole horizon scale $r_{\rm H}$. If $\lambda_c\gg r_{\rm H}$, the boson's wave function spreads out and the black hole only absorbs a very small fraction. In this case the cross-section is much smaller than the geometric area of the horizon and the accretion rate is suppressed. For very small $\mu$ (e.g.\ $\mu\sim10^{-22}$–$10^{-20}$ eV) and large SMBHs, the accretion (or absorption) rate is negligible over cosmological timescales~\cite{urena2002supermassive,annulli2020response}; however, depending on the particle mass and the black hole growth history, there can be regimes where the timescale becomes comparable to or shorter than cosmic ages.

An ultralight bosonic field with particle mass in the range $\mu\sim10^{-14}$–$10^{-19}$ eV, comprising a fraction of the total dark matter, could also accrete very slowly onto early black hole seeds (stellar-mass or intermediate mass). But as those seeds grow via ordinary accretion or mergers, there will come an epoch at which $\mu M_{\rm BH}$ crosses a threshold so that scalar field accretion (via quasi-bound state decay) becomes non‐negligible. The change in the scalar field accretion rate as the black hole mass grows was already shown in numerical simulations in~\cite{sanchis2016quasistationary} and in the catastrophic absorption of a boson star by a parasitic black hole~\cite{cardoso2022parasitic}. At that point, there is a rapid boost in the growth rate, enabling the black hole to reach supermassive masses ($10^6$–$10^9\,M_\odot$) more quickly than by baryonic/accretion/merger alone.

In what follows, using approximations adapted from the quasi-bound state calculations~\cite{barranco2011black,barranco2012schwarzschild,barranco2014schwarzschild}, we integrate the accretion rate of scalar field dark matter over cosmological time as a function of $\mu$ and black hole mass. We demonstrate that this \emph{dark-to-black} conversion leads to exponential black hole growth. Modest seeds in the range $10^2$--$10^5 M_\odot$ embedded in bosonic clouds of $\sim 10^6-10^8 M_\odot$ can reach supermassive scales within $\sim 10^8$ yrs. This process operates independently of baryonic accretion, providing a natural and testable mechanism to explain the emergence of high-redshift SMBHs and linking early-universe black hole demographics to fundamental dark matter physics.

\section{Setup}
We analyze the accretion of massive scalar fields into black holes within the framework of general relativity within Einstein's equations
$R_{\alpha\beta}-\frac{1}{2}Rg_{\alpha\beta} = 8\pi T_{\alpha\beta}$, where $R_{\alpha\beta}$ and $R$ are the Ricci tensor and Ricci scalar associated to the metric $g_{\alpha\beta}$. To isolate the essential features of the accretion process, we work in the test-field limit, neglecting the backreaction of the scalar field on the spacetime geometry. The background is assumed to be spherically symmetric, with the line element taken as $ds^2 = -f(r)dt^2+\frac{1}{f(r)}dr^2 + r^2 \left(d\theta^2+\sin^2\theta d\varphi^2\right),$
where $f(r) = 1-\frac{r_{H}}{r}$ with $r_{H}=2M_{\rm BH}$. 
Conservation of the stress energy tensor implies that the scalar field obeys the 
Klein Gordon equation $\nabla_\alpha\nabla^\alpha \Psi=\mu^2\Psi$.
We focus on quasi-stationary states, for which the scalar field has the form $\Psi=e^{-i\sigma t}\psi(r)$.
Furthermore, we are interested in solutions of the Klein–Gordon equation, subject to ingoing boundary conditions at the horizon and outgoing behavior at spatial infinity. The spectral properties of this system have been extensively studied~\cite{Damour:1976kh,Zouros:1979iw,Detweiler:1980uk,Furuhashi:2004jk}, and the behavior of scalar fields in the non-rotating case has been analyzed in detail~\cite{barranco2011black,barranco2012schwarzschild,barranco2014schwarzschild,Barranco:2013rua}. 
In the regime $\mu M_{\rm BH}\ll1$, it has been shown that the dynamical accretion of generic scalar field configurations onto a Schwarzschild black hole is well described by the quasi--bound state spectrum of a massive scalar field~\cite{barranco2012schwarzschild,Baumann:2019eav}, concluding that such spectrum closely resembles that of the hydrogen atom, with a complex frequency  $\sigma=\omega+i\gamma$, whose real ($\omega$)  and imaginary ($\gamma$) parts are given, for a spherically symmetric distribution, by
\begin{eqnarray}
M_{\rm BH}\omega &=& \mu M_{\rm BH}\left[1-(\mu M_{\rm BH})^2 \right]^{1/2}, \nonumber\\
M_{\rm BH}\gamma &=& -8(\mu M_{\rm BH})^{6},
\label{eq:freqs}
\end{eqnarray}
where the imaginary part $\gamma$ encodes the absorption of scalar field at the black hole horizon. Although these expressions are derived in the test-field approximation, we use them to model a slow, gradual mass transfer from the scalar cloud to the black hole, in which the accreted mass induces a small incremental change in $M_{\rm BH}$ and a corresponding depletion of the cloud, in a cyclic process. This approximation is supported by fully relativistic numerical simulations~\cite{cardoso2022parasitic}. 
As shown in Refs.~\cite{detweiler1980klein,Baumann:2019eav}, the dominant contribution arises from the ground state ($n=0$ in our case), while higher-$n$ states are subdominant.

\begin{figure}[t!]
\begin{center}
\includegraphics[width=0.5\textwidth]{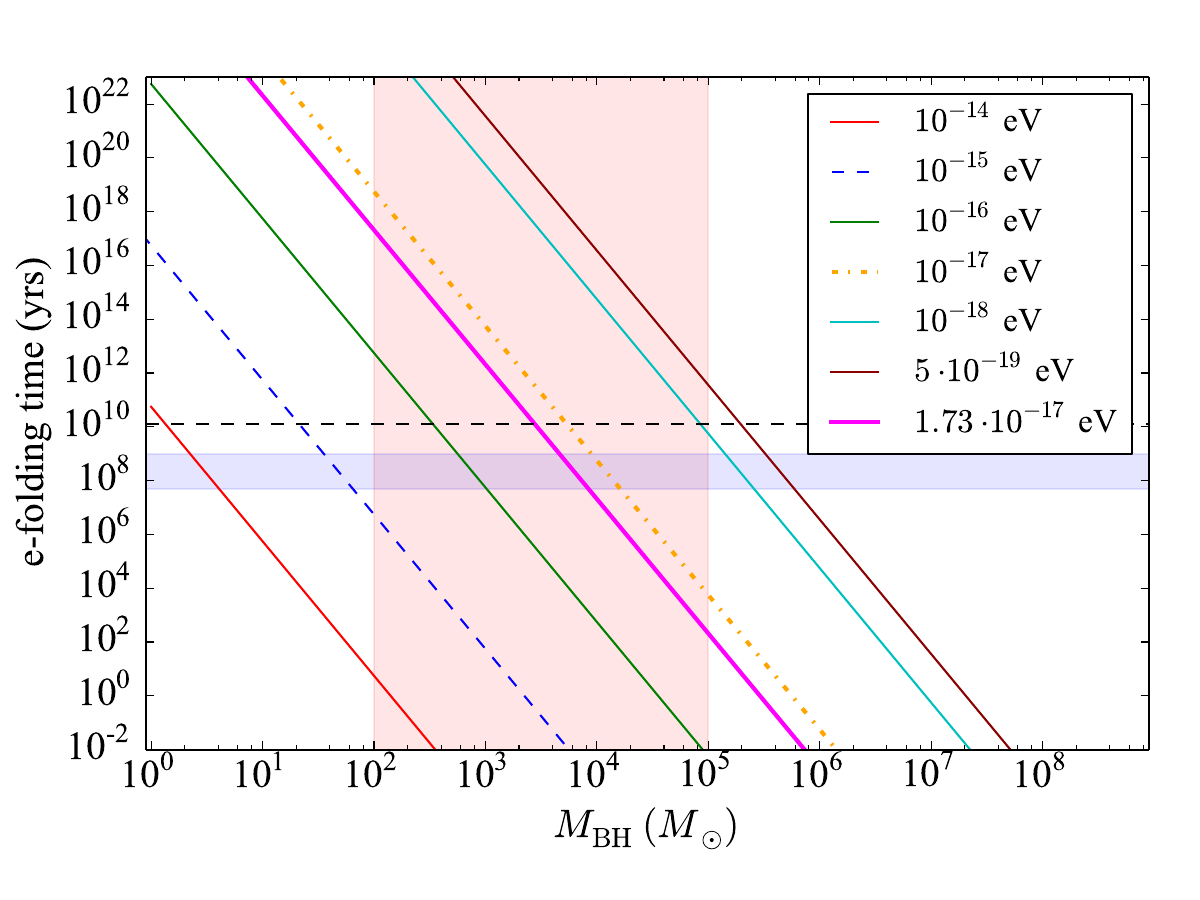}\vspace{-0.3cm}
\includegraphics[width=0.51\textwidth]{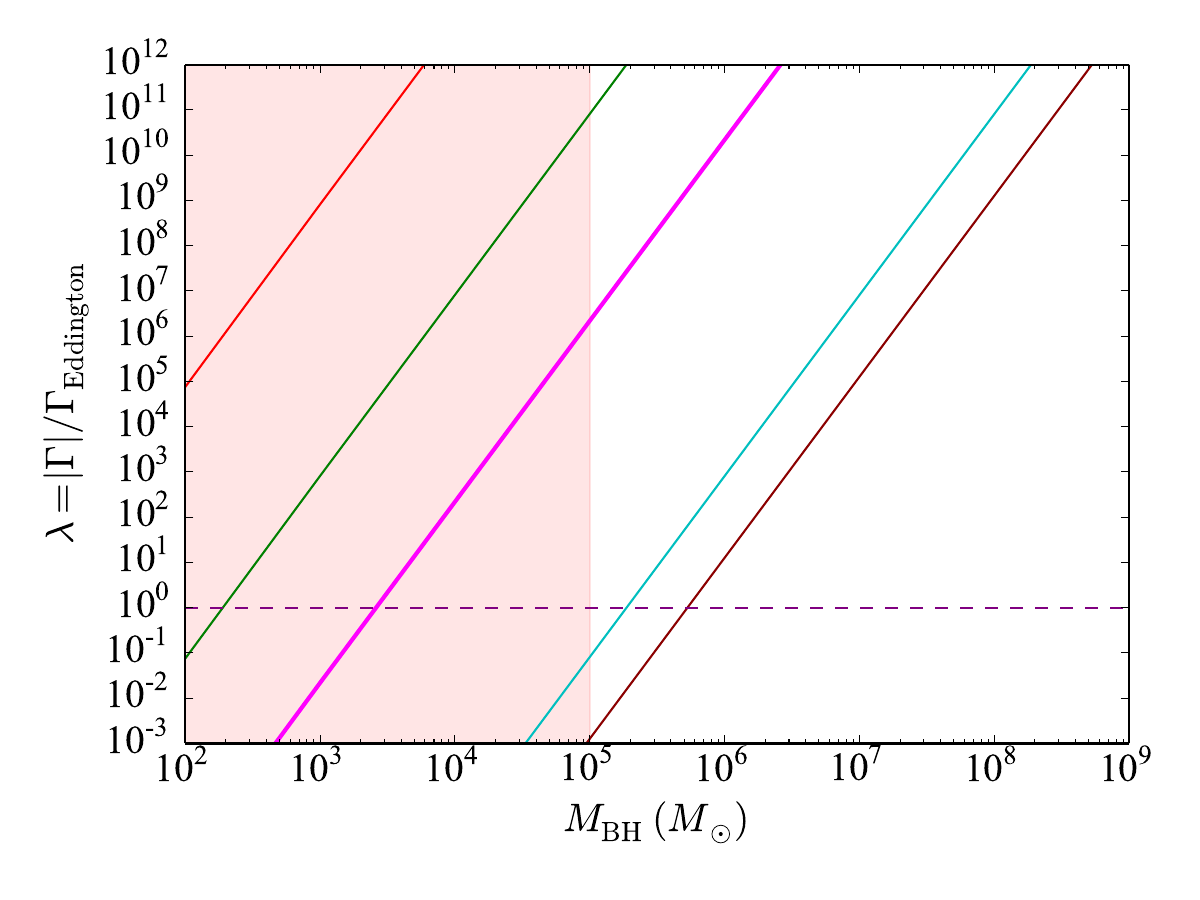}
\vspace{-0.5cm}\\
\caption{Top panel: 
Characteristic timescale (e-folding time) in years for different scalar field particle masses $\mu$ as a function of the black hole mass. Bottom panel: ratio $\lambda=|\Gamma|/\Gamma_{\rm Eddington}$ between the scalar field and Eddington accretion rates.}
\label{fig1}
\end{center}
\end{figure}

The decay of the scalar field amplitude is characterized by $\gamma$, but we will instead consider the decay of the scalar field energy given by
\begin{equation}
\Gamma = 2\gamma=-16\mu^6M_{\rm BH}^5.
\end{equation}

In the top panel of Fig.~\ref{fig1} we show the e-folding timescale of the quasi-bound state, defined as $t_E = |\Gamma|^{-1}\propto\mu^6M_{\rm BH}^{5}$ from Eq.~(\ref{eq:freqs}), as a function of the black hole mass for several representative boson masses $\mu$. For reference, the age of the Universe ($t_{\rm U}\simeq1.3\times10^{10}\,\mathrm{yr}$) is indicated by the black dashed line: above this level, the cloud is effectively stable on cosmological timescales. The red shaded region marks the expected range of SMBH seed masses, while the blue band highlights $t_E=0.05$--$1$~Gyr, the interval required for rapid SMBH growth at high redshift. Depending on potential new observations at even high redshift, this time interval could be extended. At first sight, the overlap of these bands appears to define the viable boson mass range. However, this conclusion holds only if the scalar cloud is non-self-gravitating and the black hole mass remains fixed. In the following we relax these assumptions and show that accounting for black hole growth qualitatively reshapes the allowed parameter space. The bottom panel compares the scalar accretion rate with the Eddington limit, $\lambda=|\Gamma|/\Gamma_{\rm Eddington}$, where $\Gamma_{\rm Eddington}=2.2\times10^{-8}(M_{\rm BH}/M_{\odot})\,[M_{\odot}\,\mathrm{yr}^{-1}]$ for maximal radiative efficiency. Once a critical mass is exceeded, scalar accretion can surpass the Eddington rate by orders of magnitude, since it is exponential but mass-dependent, unlike universal Eddington growth.

\begin{figure}[t!]
\begin{center}
\includegraphics[width=0.5\textwidth]{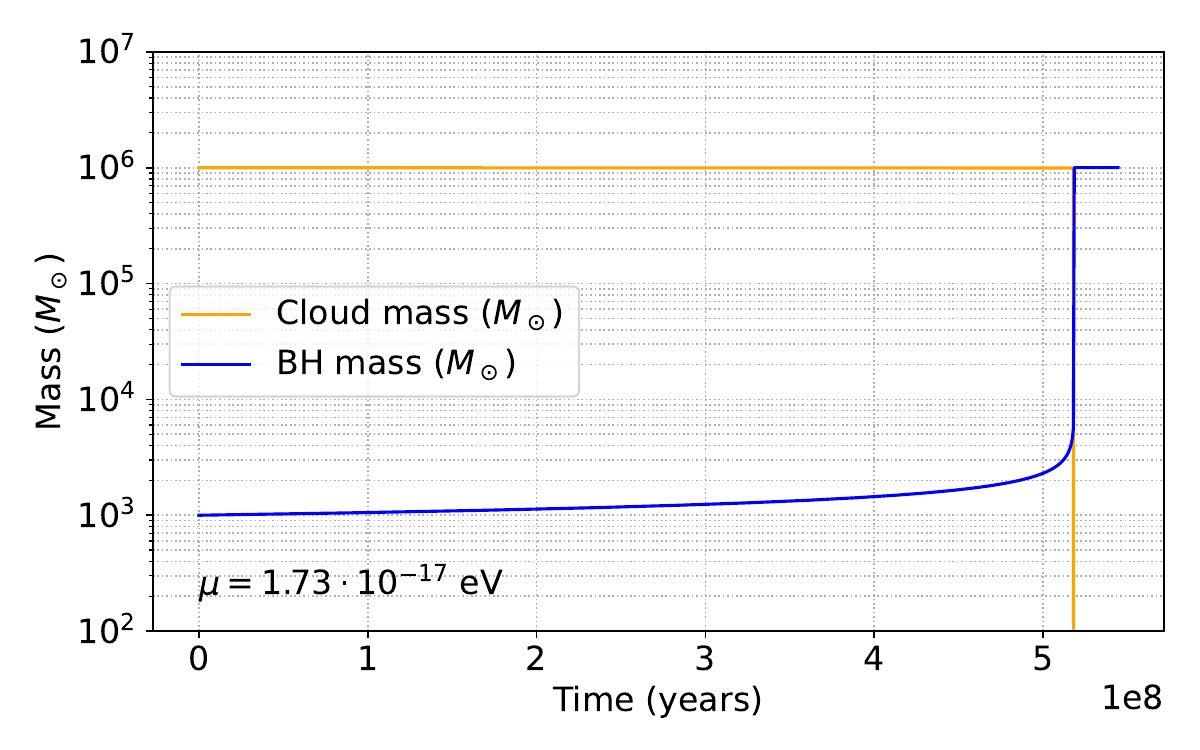}
\includegraphics[width=0.5\textwidth]{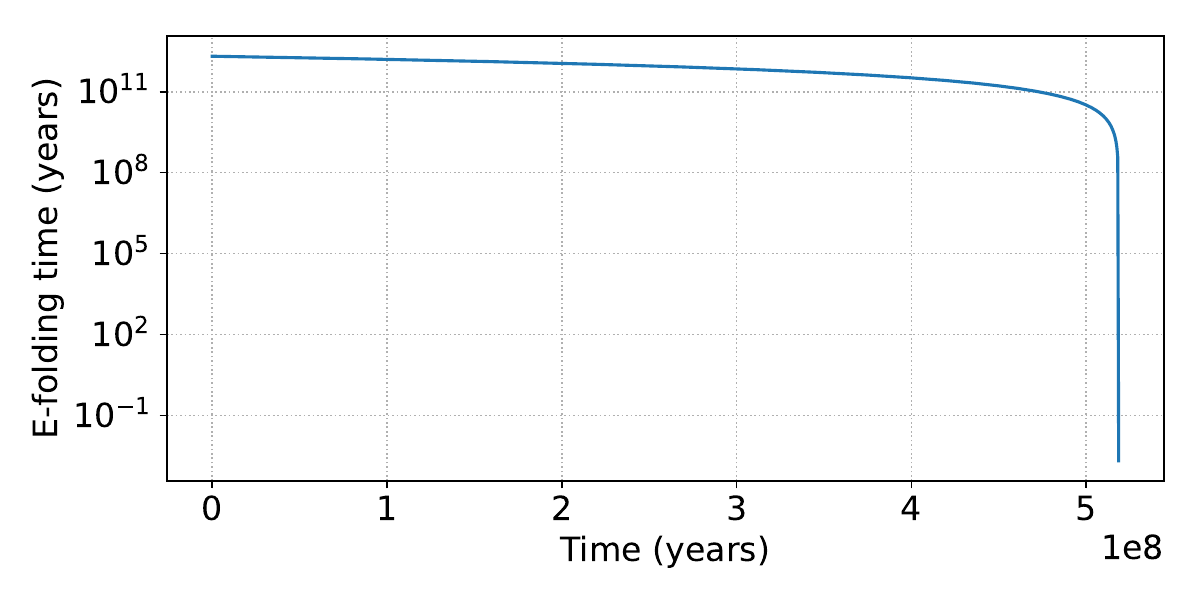}
\vspace{-0.5cm}\\
\caption{Top panel: Black hole mass as a function of the accretion time for $\mu=1.73\times10^{-17}$ eV. Bottom panel: Same for the e-folding time in years.}
\label{fig2}
\end{center}
\end{figure}

\section{Accretion of matter and black hole growth.} To model the mass growth of the black hole, we adopt an adiabatic approximation in which the scalar field is gradually absorbed by the horizon, leading to a corresponding increase in the black hole mass. The black hole evolves smoothly, while at each instant the scalar field can be regarded as approximately stationary with respect to the background geometry.  The black hole mass $M_{\rm BH}(t)$ grows at the expense of the cloud mass $M_c(t)$ according to
\begin{equation}\label{eq:growth}
\dot M_{\rm BH} = |\Gamma| M_c, 
\qquad
\dot M_c = -|\Gamma| M_c ,
\end{equation}
where the accretion rate is
given by Eq.~\eqref{eq:freqs}.
The equations conserve the total mass $M_{\rm BH}+M_c$ and yield in the late stages an exponential depletion of the cloud. We assume that the cloud loses energy solely via absorption at the black hole horizon. In exact spherical symmetry there is no gravitational radiation, and scalar radiation is strongly suppressed during the rapid accretion phase, at least for complex scalar fields, as shown by fully relativistic simulations~\cite{cardoso2022parasitic}. Relaxing spherical symmetry would generically allow additional loss channels and alter the effective accretion rate.

In our scenario, the scalar cloud represents a dense dark matter component surrounding the black hole. In fuzzy (ultralight scalar) dark matter cosmologies, solitonic cores naturally arise from gravitational collapse of small initial density fluctuations, forming on roughly the halo free-fall timescale, $t_{\rm cond}\sim 10^{8}$--$10^{9}\,$yrs for particle masses $\mu\sim 10^{-22}$--$10^{-20}\,$eV~\cite{schive2014cosmic,hui2017ultralight,davies2020fuzzy,hui2021wave}. Simulations show such cores exist at high redshift ($z\gtrsim7$) with masses $M\gtrsim 10^9\,M_\odot$~\cite{chiu2025boosting}, making a massive scalar cloud around an early black hole seed astrophysically plausible. For the scalar masses considered here ($\mu\sim10^{-19}$–$10^{-15}\,\mathrm{eV}$), the cloud should not be directly interpreted as a kiloparsec-scale halo or solitonic core. Instead, it can be modeled as a compact bosonic cloud overdensity formed through gravitational cooling~\cite{seidel1994formation,liebling2023dynamical,di2018dynamical}, as commonly found in the nonlinear evolution of self-gravitating scalar fields. Such configurations have a characteristic size $R\sim\mu^{-1}$ and form on dynamical timescales $t\sim R\sim\mu^{-1}$, which can be potentially much shorter than the initial epoch considered in our analysis ($t_i\sim10^{9}\,\mathrm{yrs}$). In this work, we therefore treat the scalar cloud as a pre-existing reservoir of dark matter surrounding the black hole and focus on its subsequent accretion.

We solve the system using a high accuracy explicit Runge–Kutta method (5th order with embedded 4th order) with adaptive step size, terminating the integration when the cloud mass falls below $10^{-15} M_\odot$. We extract the instantaneous accretion rate $\dot M_{\rm BH}$ in $M_\odot \, {\rm yr}^{-1}$.

\begin{figure}[t!]
\begin{center}
\includegraphics[width=0.5\textwidth]{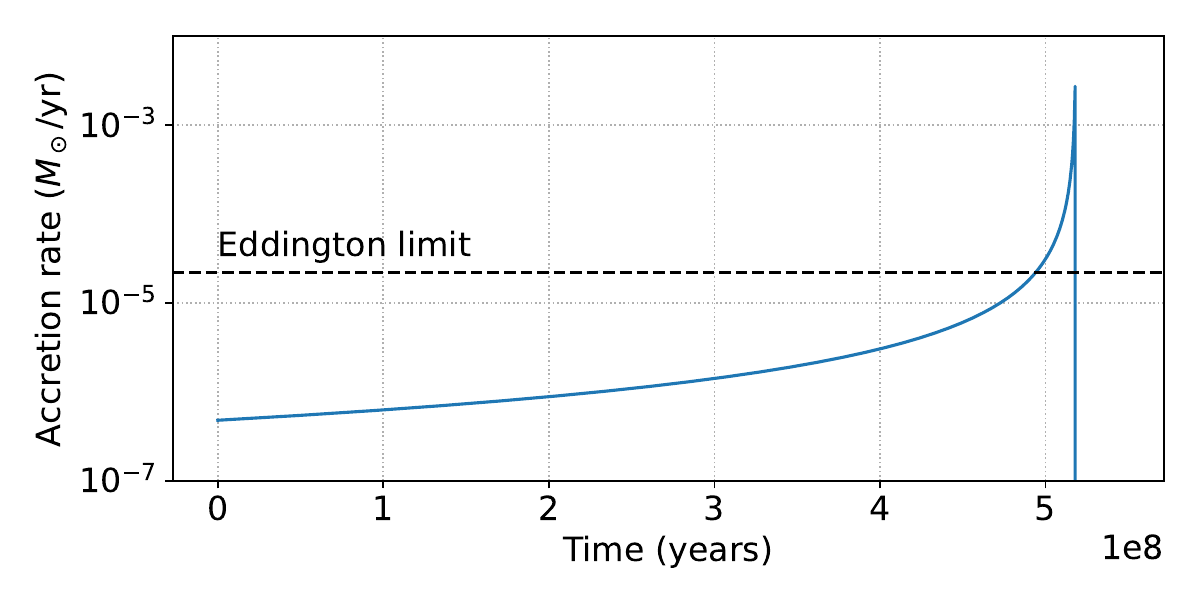}
\includegraphics[width=0.5\textwidth]{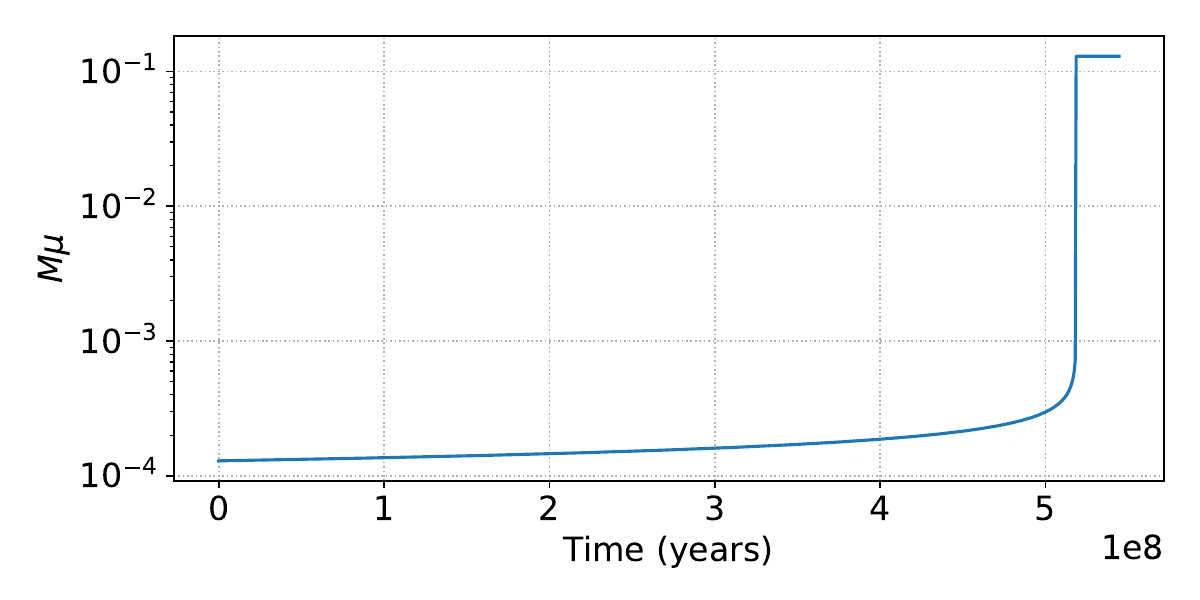}
\vspace{-0.5cm}\\
\caption{Same as Fig.~\ref{fig2} for the accretion rate (top panel) and the parameter $\mu M_{\rm BH}$ (bottom panel).}
\label{fig3}
\end{center}
\end{figure}

As an illustrative example, we consider the initial parameters $\mu = 1.73\times10^{-17}\,\mathrm{eV}$, a boson cloud mass of $M_{c} = 10^{6}\,M_{\odot}$, and a black hole seed of $M_{\rm BH}^{\rm seed}$ := $M_{\rm BH}(t=0) = 10^{3}\,M_{\odot}$. The initial configuration corresponds to $\mu M_{\rm BH} \sim 10^{-4}$ and an e-folding time of $t_{E} \simeq 1.4\times10^{12}\,\mathrm{yrs}$, much longer than the age of the Universe. Solving Eqs.~(\ref{eq:growth}), we obtain the time evolution shown in Figs.~\ref{fig2} and \ref{fig3}, where we track the black hole mass, boson cloud mass, e-folding time, the dimensionless parameter $\mu M_{\rm BH}$, and the accretion rate. The results show that the black hole grows extremely slowly at first: after more than $10^{8}$\,yrs, its mass has only doubled to $M_{\rm BH}\simeq 2\times10^{3}\,M_{\odot}$. Once the system approaches the resonant regime, however, the growth accelerates sharply, with the mass increasing from $2\times10^3$ to $4\times10^3\,M_{\odot}$ in just $\sim 10^{7}$\,yrs. The subsequent depletion of the boson cloud proceeds in a runaway fashion, completing almost instantaneously on cosmological scales, within $\sim 10^{6}$\,yrs. After that, in the span of only half a million years, the black hole attains $10^6\,M_{\odot}$. This runaway behavior is driven by the subsequent decrease in the e-folding time and the rise in the accretion rate (see~\cite{cardoso2022parasitic}). As shown in the bottom panel of Fig.~\ref{fig2}, doubling the black hole mass reduces $t_{E}$ by roughly two orders of magnitude, and another order of magnitude is lost with the next doubling. By the time the black hole reaches $M_{\rm BH}\sim 4000\,M_{\odot}$, the e-folding time becomes shorter than the overall evolutionary timescale ($\sim 10^{8}$\,yrs), and when $M_{\rm BH}\sim 10^4\,M_{\odot}$ the e-folding time is $t_{\rm E}\simeq1.4\times10^{7}$\,yrs, leading to an effectively exponential depletion of the boson cloud in about $10^5$ yrs.

In Fig.~\ref{fig3} we present the evolution of the accretion rate $\dot M_{\rm BH}$ and of the dimensionless parameter $\mu M_{\rm BH}$, offering a different perspective on the same dynamics. As the system evolves, $\mu M_{\rm BH}$ increases steadily and saturates at a final value of $\sim 0.13$. In the top panel of Fig.~\ref{fig3}, the black dashed horizontal line marks the Eddington accretion limit for the chosen black hole seed mass. Therefore, the accretion rate becomes super-Eddington for a short period of time. This illustrates how the black hole mass growth naturally drives the system into the resonant regime where accretion becomes most efficient in the expected timescale. Taken together with the runaway behavior discussed above, these results demonstrate a possible pathway for black holes to experience accelerated growth phases, providing a natural mechanism to account for the emergence of supermassive black holes already observed at high redshift.

\begin{table}[t!]
\centering
\caption{Illustrative depletion timescales for different values of $\mu$ and initial parameters: black hole seed mass and initial boson cloud mass.}
\begin{tabular}{|c|c|c|c|}
\hline
$\mu$ [eV]&$M_{\rm BH}^{\rm seed}[M_{\odot}]$&$M_c\,[M_{\odot}]$&Depletion timescale [yrs]\\ 
\hline
$1.00\cdot10^{-16}$&$1\times10^2$&$10^6$&$1.4\times10^{8}$\\
$1.73\cdot10^{-17}$&$5\times10^2$&$10^6$&$8.3\times10^{9}$\\
$1.73\cdot10^{-17}$&$1\times10^3$&$10^6$&$5.2\times10^{8}$\\
\hline
$5.00\cdot10^{-18}$&$5\times10^3$&$10^7$&$1.4\times10^{8}$\\
$7.00\cdot10^{-19}$&$5\times10^3$&$10^8$&$1.9\times10^{12}$\\
$7.00\cdot10^{-19}$&$5\times10^4$&$10^8$&$1.9\times10^{8}$\\
\hline
$5.00\cdot10^{-19}$&$1\times10^5$&$10^7$&$8.9\times10^{8}$\\
$5.00\cdot10^{-19}$&$6\times10^4$&$10^8$&$6.9\times10^{8}$\\
$5.00\cdot10^{-19}$&$1\times10^5$&$10^8$&$8.9\times10^{7}$\\
$3.00\cdot10^{-19}$&$8\times10^4$&$10^9$&$4.7\times10^{8}$\\
\hline
$9.00\cdot10^{-20}$&$5\times10^5$&$10^9$&$4.2\times10^{8}$\\
\hline
\end{tabular}
\label{table1}
\end{table}

\begin{figure}[t!]
\begin{center}
\includegraphics[width=0.51\textwidth]{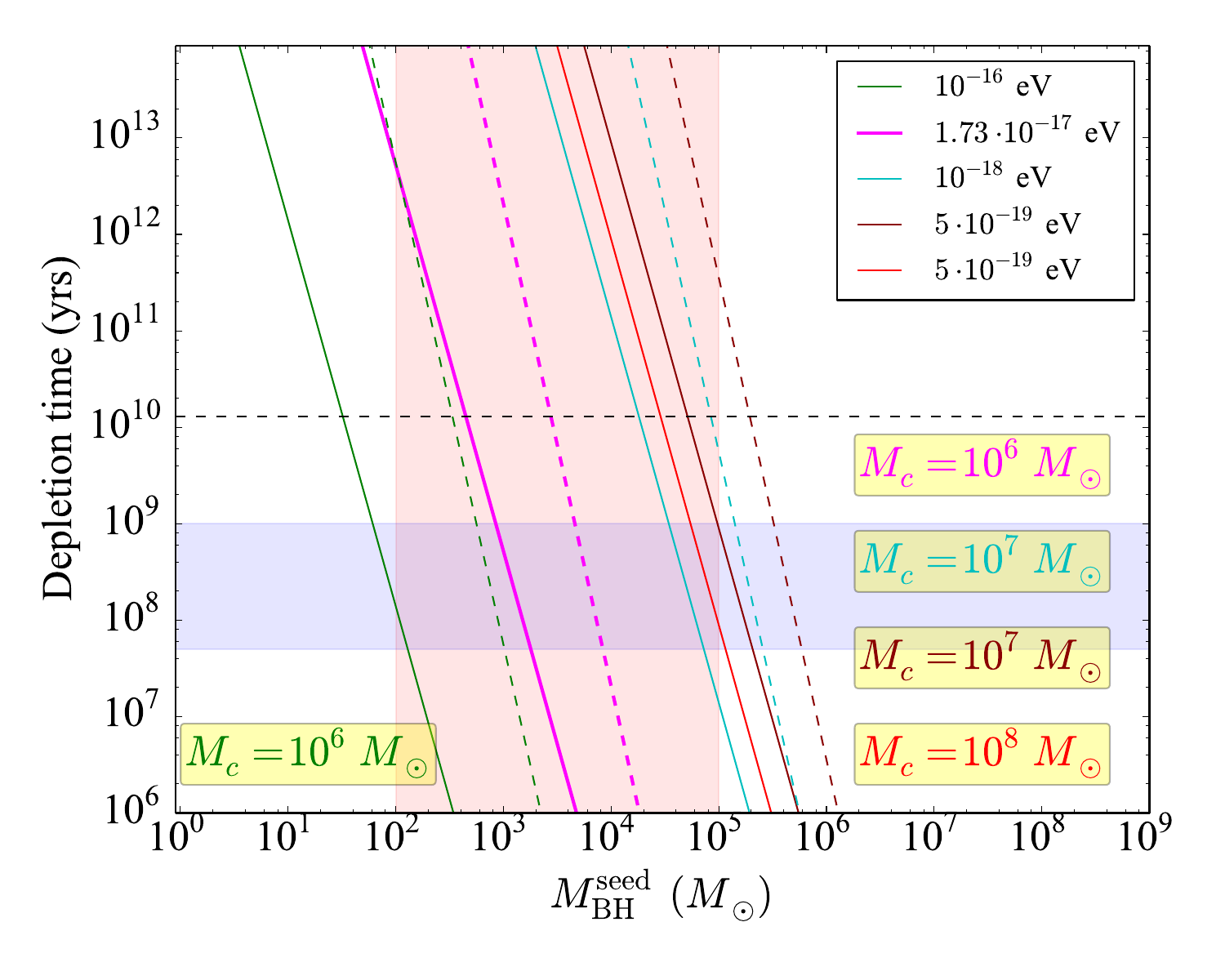}
\vspace{-0.5cm}\\
\caption{Depletion time (solid line) and e-folding time (dashed lines) as a function of the initial black hole seed mass $M^{\rm seed}_{\rm BH}$  for different boson masses $\mu$ and initial boson cloud masses $M_c$.  Increasing $M_c$ reduces the total absorption time at fixed $\mu$ by accelerating the early growth of the black hole.}
\label{fig4}
\end{center}
\end{figure}

In Table~\ref{table1} we summarize characteristic depletion timescales for several illustrative cases. 
The overall trends are consistent with expectations: smaller boson masses $\mu$ yield longer depletion times, 
whereas more massive black hole seeds shorten them. Increasing the initial boson cloud mass enhances the 
efficiency of absorption at fixed $\mu$ and $M_{\rm BH}$, since the black hole grows more rapidly from the outset. 
Figure~\ref{fig4} shows the total time required for a black hole to absorb its 
surrounding cloud. Compared with Fig.~\ref{fig1}, the slope of the curves is altered: lighter bosonic fields, 
which previously appeared inefficient, now fall within the range compatible with astrophysically plausible 
black hole seed masses and cosmological growth times. This highlights how the interplay between boson mass, black hole seed mass, 
and cloud mass jointly determines whether \emph{dark-to-black} conversion can account for early SMBH growth. Light bosons in the range $\mu \in [10^{-17},\,10^{-16}]~\mathrm{eV}$ require relatively small black hole seeds, $M_{\rm BH}^{\rm seed} \in [10^2,\,10^3]\,M_{\odot}$, which can grow to  $10^6\,M_{\odot}$ within a few hundred million years. In contrast,  more massive SMBHs in the range  $10^{7}$--$10^{9}\,M_{\odot}$ demands lighter bosons, $\mu \in [10^{-19},\,10^{-18}]~\mathrm{eV}$, together with substantially larger seeds of order $10^{4}$--$10^{5}\,M_{\odot}$.

\section{Discussion and Outlook} The process described here can be thought of as a \emph{dark-to-black} conversion: dark matter becomes part of the black holes, losing casual contact with the rest of the Universe. This channel would have been particularly efficient in the early Universe, when proto-galaxies were strongly dark matter dominated and baryonic collapse was still inefficient. The selective nature of the resonance condition ($\mu M_{\rm BH} \sim 1$) implies that only a fraction of the scalar field component participates in the process, thus avoiding tensions with large-scale structure and Lyman-$\alpha$ forest constraints~\cite{irvsivc2017first,armengaud2017constraining}, while still producing a dynamically relevant effect, in good agreement with recent studies in the full non-linear regime~\cite{cardoso2022parasitic}. This efficient transfer would simultaneously accelerate the growth of SMBHs and reduce the central dark matter density in their vicinity, deepening the local potential well and enhancing subsequent baryonic inflows. Its observational imprints would be indirect, manifesting deviations from standard halo profiles, lensing signatures, or SMBH–host scaling relations at high redshift. In this way, \emph{dark-to-black} conversion could represent a key missing link between dark matter physics and the unexpectedly rapid emergence of massive black holes in the early Universe.

In this sense, our proposal complements, rather than replaces, existing baryonic channels for early black hole growth. Although super-Eddington accretion, mergers of stellar-mass black holes, and the collapse of supermassive Pop~III stars provide plausible growth routes, each of these mechanisms faces significant theoretical and observational challenges at $z \gtrsim 10$~\cite{chen2014general,shi2023hyper,nagele2024formation,gaete2024supermassive}. In contrast, the \emph{dark-to-black} mechanism operates independently of the baryonic supply, allowing rapid growth once the resonance condition $\mu M_{\rm BH} \lesssim 1$ is reached. Indeed, accretion remains suppressed as long as the Compton wavelength $\lambda_c$ of the field exceeds the black hole size $r_{   H}$. Once the two scales become comparable, absorption becomes efficient and leads to rapid black hole growth. In this picture, black holes act as selective sinks for a fraction of the dark matter component, while baryonic accretion continues in parallel. 

This framework can be naturally generalized to incorporate additional physical processes, such as the combined
accretion of baryonic matter or possible direct interactions between the scalar field and baryonic components, 
thereby providing a more comprehensive description of black hole formation and evolution.
The combination of these processes could naturally account for the existence of SMBHs with $M_{\rm BH} \sim 10^8-10^9\,M_\odot$ at redshifts $z \gtrsim 15$, reconciling recent JWST observations with cosmological timescales. Moreover, the efficiency of the \emph{dark-to-black} mechanism is controlled by fundamental particle physics parameters, opening a new observational window into the microphysical nature of dark matter through high-redshift black hole demographics.

Because the resonance is sensitive to $M_{\rm BH}$ and $\mu$ and the cloud supplies are finite, this channel is effectively a one-time conversion per halo, that is, once the local coherent reservoir is depleted by accretion, the same galaxy cannot generically repeat the event for smaller black holes formed later. Any remnant cloud would be small and the process leaves only gravitational imprints (modified central halo profiles, lensing anomalies, or altered SMBH–host scaling) or indirect probes via superradiance/spin limits in the appropriate mass windows~\cite{cardoso2018constraining,cunha2019eht}. Although we have focused on minimally coupled scalar fields, analogous quasibound phenomena exist for other bosonic fields~\cite{rosa2012massive}, massive Dirac fields~\cite{lasenby2005bound,dolan2015bound} and for mixed boson–fermion couplings. Recent works have revisited fermionic quasibound spectra around Schwarzschild/Kerr black holes and explored boson-driven fermion production~\cite{chen2025revisiting,chen2025black}, suggesting that the qualitative resonant absorption channel may extend beyond scalars to a broader class of dark-sector candidates.

\textit{\textbf{Acknowledgements}}. NSG acknowledges support from the Spanish Ministry of Science, Innovation, and Universities via the Ram\'on y Cajal programme (grant RYC2022-037424-I), funded by MCIN/AEI/10.13039/501100011033 and by ``ESF Investing in your future”. This work is further supported by the Spanish Agencia Estatal de Investigaci\'on (Grant PID2021-125485NB-C21 and PID2024-159689NB-C21) funded by MCIN/AEI/10.13039/501100011033 and ERDF A way of making Europe,   and by the European Horizon Europe staff exchange (SE) programme HORIZON-MSCA2021-SE-01 Grant No. NewFunFiCO-101086251. This work is also supported by CIDMA under the FCT Multi-Annual Financing Program for R\&D Units (UID/4106/2025) through the Portuguese Foundation for Science and Technology (FCT), project 2024.05617.CERN (\url{https://doi.org/10.54499/2024.05617.CERN}) and by DGAPA-UNAM through grant IN110523.  

\bibliography{ref}

\end{document}